\documentclass[prd,reprint,nofootinbib,superscriptaddress]{revtex4-2}
\usepackage{bm}
\usepackage{amsmath}
\usepackage{amssymb}
\usepackage{graphicx}
\usepackage{float}
\usepackage{subfigure}
\usepackage{hyperref}
\usepackage{CJK}
\hypersetup{
colorlinks=true,
linkcolor=red,
citecolor=blue,
}

\begin{document}


\title{Source localizations with the network of space-based gravitational wave detectors}

\author{Chunyu Zhang}
\email{chunyuzhang@hust.edu.cn}
\affiliation{School of Physics, Huazhong University of Science and Technology,
Wuhan, Hubei 430074, China}

\author{Yungui Gong}
\email{Corresponding author. yggong@hust.edu.cn}
\affiliation{School of Physics, Huazhong University of Science and Technology, Wuhan, Hubei 430074, China}

\author{Chao Zhang}
\email{chao\_zhang@hust.edu.cn}
\affiliation{School of Physics, Huazhong University of Science and Technology,
Wuhan, Hubei 430074, China}

\begin{abstract}
The sky localization of the gravitational wave (GW) source is an important scientific objective for GW observations.
A network of space-based GW detectors dramatically improves the sky localization accuracy compared with an individual detector not only in the inspiral stage but also in the ringdown stage.
It is interesting to explore what plays an important role in the improvement.
We find that the angle between the detector planes dominates the improvement, and the time delay is the next important factor.
A detector network can dramatically improve the source localization for short signals and long signals with most contributions to the signal-to-noise ratio (SNR) coming from a small part of the signal in a short time,
and the more SNR contributed by smaller parts, the better improvement by the network.
We also find the effects of the arm length in the transfer function and higher harmonics are negligible for source localization with the detector network.

\end{abstract}

\preprint{2112.02299}

\maketitle


\section{Introduction}

The Laser Interferometer Gravitational-Wave Observatory (LIGO) Scientific Collaboration and the Virgo Collaboration have reported tens of confirmed gravitational wave (GW) detections \cite{Abbott:2016blz,Abbott:2016nmj,Abbott:2017vtc,Abbott:2017oio,TheLIGOScientific:2017qsa,Abbott:2017gyy,LIGOScientific:2018mvr,TheLIGOScientific:2016agk,Abbott:2020uma,LIGOScientific:2020stg,Abbott:2020khf,Abbott:2020tfl,LIGOScientific:2020ibl,LIGOScientific:2021djp},
which provide a new avenue to probe the nature of gravity and spacetime in the nonlinear and strong field regimes.
Ground-based GW detectors, such as Advanced LIGO \cite{Harry:2010zz,TheLIGOScientific:2014jea}, Advanced Virgo \cite{TheVirgo:2014hva} and Kamioka Gravitational Wave Detector (KAGRA) \cite{Somiya:2011np,Aso:2013eba}, operating in the $10-10^4$ Hz frequency band, normally detect stellar-mass binary mergers with small signal-to-noise ratios (SNRs).
The proposed space-based GW detectors such as Laser Interferometer Space Antenna (LISA) \cite{Danzmann:1997hm,Audley:2017drz}, TianQin \cite{Luo:2015ght}, and Taiji \cite{Hu:2017mde} probe GWs in the millihertz frequency band, while Deci-hertz Interferometer Gravitational Wave Observatory (DECIGO) \cite{Kawamura:2011zz} operates in the 0.1 to 10 Hz frequency band.
Thus, space-based GW detectors can detect GWs from massive black hole (BH) binary mergers with large SNRs,
and the detected signals can be well used to probe the nature of BHs, localize sources, estimate their parameters, etc.
In particular, localizing the sky position of the GW source is a key scientific goal for GW observations.
An accurate source position is essential for the follow-up observations of electromagnetic counterparts or the statistical identification of the host galaxy if no counterpart is present. 
Furthermore, we can study the thermal history of the Universe and measure cosmological parameters with GW observations.
For example, GWs with accurate information about the source positions can be regarded as standard sirens \cite{Schutz:1986gp, Holz:2005df} to understand the problem of Hubble tension \cite{Riess:2019cxk}.

GWs from compact binary coalescences are described by inspiral, merger, and ringdown phases, with increasing frequency.
The inspiral waves, at the stage of orbiting until the innermost stable orbit, can be analyzed by
the post-Newtonian theory \cite{Peters:1963ux,Peters:1964zz}, BH perturbation theory \cite{Teukolsky:1973ha,Poisson:1995vs}, etc.
The merger waveform is normally described by phenomenological models, effective-one-body models calibrated to numerical-relativity simulations, etc \cite{Husa:2015iqa,London:2017bcn,Pratten:2020fqn,Garcia-Quiros:2020qpx,Buonanno:1998gg,Buonanno:2000ef,Yunes:2009ef,Bohe:2016gbl,Cotesta:2020qhw,Blackman:2017pcm,Mroue:2013xna,Boyle:2019kee}.
The ringdown signal originating from the distorted final BH, comprises a superposition of quasinormal modes (QNMs) \cite{Leaver:1985ax,Berti:2005ys}.
Each mode has a complex frequency, the real part is the oscillation frequency, and the imaginary part is the inverse of the damping time.
These frequencies are determined by the mass and angular momentum of the final BH, and the amplitude and phase of each mode are determined by the specific process when the final BH forms.
Ground-based GW observatories detect GW signals lasting within a few seconds to minutes since GW signals are inferior to the noise until the later time of the inspiral phase,
so at least three ground-based observatories at widely separated sites are required to localize compact binaries \cite{Fairhurst:2010is,Grover:2013sha}.
However, space-based GW detectors can measure GWs lasting from months to years due to the large GW amplitude and the low orbital frequency of massive BH binaries.
The early inspiral waves are always regarded as monochromatic waves due to their slow evolution.
For monochromatic waves, space-based GW detectors use the modulations of the amplitude and phase caused by the motion around the Sun to localize the source \cite{Cutler:1997ta,Cutler:1998muh,Moore:1999zw,Barack:2003fp,Zhang:2020hyx,Zhang:2020drf}.
The measurement accuracy of LISA for monochromatic sources increases a few times for the observation time $T_{\rm obs}\ge2$ yr compared with that for $T_{\rm obs}=1$ yr \cite{Takahashi:2002ky}.
The transfer function can increase the measurement accuracy for monochromatic sources with $0.01\ {\rm Hz}\le f\le0.03\ {\rm Hz}$ by a few times compared with the long-wave approximation \cite{Vecchio:2004ec}.
For coalescing binaries with the total mass $\sim10^5\ M_\odot$, the transfer function can help LISA localize their sky positions ten times more accurately compared with the long-wave approximation \cite{Seto:2002uj}.

LISA and Taiji are composed of a triangle formed by three spacecrafts in a heliocentric orbit behind or ahead of the Earth by about $20^\circ$.
The angular resolution of the LISA-Taiji network depends on the configuration angle, and the network is expected to improve the sky localization of GW sources over 2 orders of magnitude than individual LISA or Taiji detector \cite{Ruan:2020smc}.
On the other hand, the LISA-Taiji network can constrain the Hubble parameter within 1\% accuracy \cite{Wang:2020dkc}.
Different from LISA and Taiji, TianQin is a geocentric detector orbiting the Earth and further rotating around the Sun together with the Earth, whose detector plane points to the source RX J0806.3+1527.
The LISA-TianQin network can improve the sky localization of Galactic double white dwarf binaries up to 3 orders of magnitude \cite{Huang:2020rjf}, if compared with a single TianQin observation.
Although the separation between LISA and TianQin is not as large as that between LISA and Taiji, the LISA-TianQin network still shows its strong ability in improving the source localization.
Moreover, the Taiji-TianQin network improves the localization of coalescence sources by two orders of magnitude compared with an individual detector \cite{Gong:2021gvw}.
Without considering the time delay, the LISA-TianQin network, Taiji-TianQin network, and LISA-Taiji network all can improve the source localization by two orders of magnitude compared with an individual detector in the ringdown stage \cite{Zhang:2021kkh}.
It seems that the time delay caused by the separation is not the primary factor in the improvement to the source localization.
Since the network of space-based GW detectors significantly improves the angular resolution, it is natural to explore what dominates the improvement and what configuration setting maximizes the improvement.
To give a robust estimation of the sky localization of the source, we employ the Fisher information matrix approximation (FIM), which is widely used to perform parameter estimation for space-based GW detectors \cite{Peterseim:1996cw,Peterseim:1997ic,Cutler:1997ta,Cutler:1998muh,Moore:1999zw,Barack:2003fp,Blaut:2011zz,Vallisneri:2007ev,Wen:2010cr,KAGRA:2013rdx,Grover:2013sha,Berry:2014jja,Singer:2015ema,Becsy:2016ofp,Zhao:2017cbb,Mills:2017urp,Fairhurst:2017mvj,Fujii:2019hdi,Ruan:2019tje,Ruan:2020smc,Feng:2019wgq,Wang:2020vkg,Huang:2020rjf,Zhang:2020hyx,Zhang:2020drf,Shuman:2021ruh,Mangiagli:2020rwz,Baibhav:2020tma,Zhang:2021kkh}.

The paper is organized as follows.
In Sec. \ref{SecMethod}, we introduce the signal in the detector and the FIM method.
In Sec. \ref{SecLocalization}, we analyze the effects of the angle between the detector planes, the time delay, the transfer function, and the higher harmonics on the source localization.
We conclude this paper in Sec. \ref{SecConclusion}.
Throughout this paper, we use units in which $G=c=1$.

\section{Fisher Information Matrix Method}
\label{SecMethod}

\subsection{Polarization tensors}
In the heliocentric coordinate $\{\hat{i},\hat{j},\hat{k}\}$, we use the source position $(\theta_s,\varphi_s)$ and the polarization angle $\psi_s$ to form the GW coordinate basis vectors $\{\hat{m},\hat{n},\hat{o}\}$ as
\begin{equation}\label{EqGWHelioVectors}
	\{\hat{m},\hat{n},\hat{o}\}=\{\hat{i},\hat{j},\hat{k}\}\times R_z\left(\varphi_s-\pi\right) R_y\left(\pi-\theta_{s}\right) R_z\left(\psi_s\right),
\end{equation}
where $\hat{o}$ is the propagating direction of GWs, and $R_x$, $R_y$ and $R_z$ are Euler rotation matrices.

In general relativity, there are two polarizations $A=+,\times$.
With polarization tensors $e^A_{ij}$,
\begin{equation}\label{EqPolar}
	e^{+}_{ij}=\hat{m}_i\hat{m}_j-\hat{n}_i\hat{n}_j,\qquad  e^{\times}_{ij}=\hat{m}_i\hat{n}_j+\hat{n}_i\hat{m}_j,
\end{equation}
we can decompose GWs into two polarizations $h_{ij}=\sum_{A=+,\times} h_A e^A_{ij}$.

\subsection{The detector signal}
\label{SecDetector}

The configurations of space-based GW detectors are generally equilateral triangles.
We can model every detector of this kind as a combination of two independent LIGO-like detectors (``I" and ``II") with the opening angle $\gamma=\pi/3$.

We consider GWs from mergers of nonspinning binary BHs.
At the inspiral stage, with the stationary phase approximation (SPA) \cite{Finn:1992xs,Cutler:1994ys,Damour:2000gg}, the frequency-domain  signal in the detector is
\begin{equation}\label{EqInspSignal}
	s(f)=
	\sum_{\ell=2}^{\infty}\sum_{m=1}^{\ell}F^A
	\left(t_{\ell m},f,\theta_s,\varphi_s,\psi_s\right)h^{\ell m}_{A}(f),
\end{equation}
where $(\ell, m)$ are the harmonic mode indices,
\begin{equation}
	F^A=\left[D_u^A(t_{\ell m}) \mathcal{T}(f,\hat{u}\cdot\hat{o})-D_v^A(t_{\ell m}) \mathcal{T}(f,\hat{v}\cdot\hat{o})\right]e^{i\Phi_D}
\end{equation}
is the response function for polarization A, $\hat{u}(t)$ and $\hat{v}(t)$ are the unit vectors of the detector's two arms, $\mathcal{T}$ is the transfer function, $D^A$ is the arm scalar, $\Phi_D$ is the Doppler shift,
\begin{equation}
\begin{split}
h_+^{\ell m}(f)=&\frac{1}{2}\left[Y_{\ell, -m}+(-1)^\ell Y^*_{\ell, m}\right]h_{\ell,-m}(f),\\
h_\times^{\ell m}(f)=&\frac{i}{2}\left[Y_{\ell, -m}-(-1)^\ell Y^*_{\ell, m}\right]h_{\ell,-m}(f)
\end{split}
\end{equation}
are GW polarizations,
$Y_{\ell, m}(\iota,0)$ are spherical harmonics of spin-weight $-2$  \cite{Wiaux:2005fm},
$\iota$ is the inclination angle of the source,
$h_{\ell,-m}=A_{\ell m}(f)e^{-i\Phi_{\ell m}(f)}$ (with $m,f>0$) are spherical harmonic modes concentrated in the positive frequency domain under the convention $h(f)=\int h(t)\exp(-2\pi ift)dt$,
the amplitude $A_{\ell m}$ and the phase $\Phi_{\ell m}=2\pi f t_c+ \phi_0-\phi_{\ell m} $ are given by the inspiral part of the \textsc{imrphenomxhm} waveform model \cite{Pratten:2020fqn,Garcia-Quiros:2020qpx},
\begin{equation}\label{EqTime}
t_{\ell m}=t_c-\frac{1}{2\pi}\frac{d\phi_{\ell m}(f)}{df}
\end{equation}
is the function of frequency \cite{Katz:2020hku,Katz:2021uax} given by SPA,
$t_c$ is the coalescence time,
and $\phi_0$ is a phase shift.

The transfer function $\mathcal{T}$ is 
\begin{equation}
	\label{EqTransfer}
	\begin{split}
		&\mathcal{T}(f,\hat{u}\cdot\hat{o})\\
		&=\frac{1}{2}\left\{{\rm sinc}\left[\frac{f(1-\hat{u}\cdot\hat{o})}{2f^*}\right]\exp\left[\frac{f(3+\hat{u}\cdot\hat{o})}{2if^*}\right]\right.\\ 
		&\quad \left.+\text{sinc}\left[\frac{f(1+\hat{u}\cdot\hat{o})}{2f^*}\right]\exp\left[\frac{f(1+\hat{u}\cdot\hat{o})}{2if^*}\right]\right\},
	\end{split}
\end{equation}
where $\text{sinc}(x)=\sin x/x$, $f^*=c/(2\pi L)$ is the transfer frequency of the detector, $c$ is the speed of light, and $L$ is the arm length of the detector.
The Doppler shift is
\begin{equation}\label{doppler}
	\Phi_D(t)=2\pi fR_e\sin(\theta_s)\cos(\omega_e t-\varphi_s+\varphi_i)/c,
\end{equation}
where $R_e=1\ {\rm AU}$ is the orbital radius, $\omega_e=2\pi/T_e$ is the orbital frequency of the Earth, $T_e=1$ year is the period, and $\varphi_i$ is the ecliptic longitude of the detector at $t=0$.
The arm scalars are defined as
\begin{equation}
	D_u^A(t)=\frac{1}{2}\hat{u}^i(t) \hat{u}^j(t) e^A_{ij},\; D_v^A(t)=\frac{1}{2}\hat{v}^i(t) \hat{v}^j(t) e^A_{ij},
\end{equation}
where the polarization tensors $e^A_{ij}$ are given by Eqs. \eqref{EqGWHelioVectors} and \eqref{EqPolar}.
For the detector I, $\hat{u}(t)$ and $\hat{v}(t)$ are given by
\begin{equation}\label{EquvI}
	\begin{split}
		\hat{u}=&\cos\left(\frac{\gamma}{2}\right)\hat{x}-\sin\left(\frac{\gamma}{2}\right)\hat{y},\\
		\hat{v}=&\cos\left(\frac{\gamma}{2}\right)\hat{x}+\sin\left(\frac{\gamma}{2}\right)\hat{y}, \\
	\end{split}
\end{equation}
and for the detector II, $\hat{u}(t)$ and $\hat{v}(t)$ are given by
\begin{equation}\label{EquvII}
	\begin{split}
		\hat{u}=&\cos\left(\frac{2\pi}{3}-\frac{\gamma}{2}\right)\hat{x}+\sin\left(\frac{2\pi}{3}-\frac{\gamma}{2}\right)\hat{y},\\
		\hat{v}=&\cos\left(\frac{2\pi}{3}+\frac{\gamma}{2}\right)\hat{x}+\sin\left(\frac{2\pi}{3}+\frac{\gamma}{2}\right)\hat{y},
	\end{split}
\end{equation}
where $\hat{x}$ and $\hat{y}$ are the  basis vectors of the detector coordinate. 
For TianQin, the basis vectors of the detector coordinate are
\begin{equation}\label{EqTQBasis}
	\{\hat{x},\hat{y},\hat{z}\}=\{\hat{i},\hat{j},\hat{k}\}\times R_z\left(\varphi_{tq}-\frac{\pi}{2}\right)  R_x\left(-\theta_{tq}\right)
    R_z\left(\omega_{tq}t\right),
\end{equation}
where ($\theta_{tq} =
94.7^\circ$, $\varphi_{tq} = 120.5^\circ$) is the direction of the source RX J0806.3+1527 \cite{Israel:2002gq,Barros:2004er,Roelofs:2010uv,Esposito:2013vja,Kupfer:2018jee}, and $\omega_{tq}$ is the rotation frequency of TianQin.
For LISA, the basis vectors of the detector coordinate are
\begin{equation}\label{EqLISABasis}
	\{\hat{x},\hat{y},\hat{z}\}=\{\hat{i},\hat{j},\hat{k}\}\times R_z\left(\omega_{e}t\right)  R_x\left(-\frac{\pi}{3}\right) R_z\left(-\omega_{e}t\right).
\end{equation}

For the ringdown waves from mergers of nonspinning binary BHs, we use the analytical expression of the signal in the detector derived in Ref. \cite{Zhang:2021kkh},
and the amplitude model \cite{Baibhav:2018rfk,Baibhav:2017jhs} determined by fitting numerical relativity simulations.
Since the phase alignment between different angular
components of the radiation is limited by the theoretical understanding of the excitation and starting times of QNMs \cite{Andersson:1995zk,Berti:2006wq,Zhang:2013ksa},
the phases of QNMs $\phi_{\ell mn}$ are treated as free parameters in our analysis of ringdown signals.

\subsection{The noise curve}

In this paper, we use the noise curve \cite{Cornish:2018dyw}
\begin{equation}\label{EqPn}
	P_n(f) = \frac{S_x}{L^2}  + \frac{2[1+\cos^2(f/f^*)]S_a}{(2\pi f)^4 L^2}\left[1+ \left(\frac{0.4\ {\rm mHz}}{f}\right)^2 \right],
\end{equation}
where $S_x$ is the position noise, $S_a$ is the acceleration noise, $L$ is the arm length,
$f^*=c/(2\pi L)$ is the transfer frequency of the detector.
For LISA, $S_x = (1.5 \times 10^{-11} \ {\rm m})^2  \ {\rm Hz}^{-1}$, $S_a = (3 \times 10^{-15} \ {\rm m}\, {\rm s}^{-2})^2 \ {\rm Hz}^{-1}$, $L=2.5\times10^9$ m  and $f^*=19.09$ mHz \cite{Audley:2017drz}.
For TianQin, $S_x = (10^{-12}\ {\rm m})^2  \ {\rm Hz}^{-1}$, $S_a = (10^{-15} \ {\rm m}\, {\rm s}^{-2})^2\  {\rm Hz}^{-1}$, $L=\sqrt{3}\times10^8$ m and $f^*=0.2755$ Hz \cite{Luo:2015ght}.
For Taiji, $S_x = (8\times10^{-12} \ {\rm m})^2  \ {\rm Hz}^{-1}$, $S_a = (3\times10^{-15} \ {\rm m}\, {\rm s}^{-2})^2\ {\rm Hz}^{-1}$, $L=3\times10^9$ m and $f^*=15.90$ mHz \cite{Ruan:2020smc}.

For LISA and Taiji, we also add the confusion noise \cite{Cornish:2018dyw}
\begin{equation}\label{EqSc}
	\begin{split}
		S_c(f) = &\frac{2.7\times10^{-45} f^{- 7/3}}{1+0.6(f/0.01909 )^{2}}\, e^{-f^{0.138} - 221  f  \sin(521 f) } \\ &\times\left[1+{\rm tanh}(1680(0.00113-f))\right]  \ {\rm Hz}^{-1},
	\end{split}
\end{equation}
to the noise curve, where $f$ is normalized by 1 Hz.

\subsection{Fisher information matrix}	

For convenience, we define the inner product of two frequency-domain signals $s_1(f)$ and $s_2(f)$ as
\begin{equation}
	(s_1|s_2)=2\int_{f_{\rm in}}^{f_{\rm out}}\frac{s_1(f)s_2^*(f)+s_1^*(f)s_2(f)}{P_{n}(f)}df.
\end{equation}
The SNR $\rho$ for a signal $s(f)$ is simply defined as
\begin{equation}
	\rho^2=(s|s).
\end{equation}
For a detected source with a significant SNR, we can use the FIM method to estimate its parameters, which is defined as
\begin{equation}
	\label{EqGammaij}
	\begin{split}
		\Gamma_{ij}=&\left(\frac{\partial s(f)}{\partial {\bm \xi}_i}\left|\frac{\partial s(f)}{\partial {\bm \xi}_j}\right.\right),
	\end{split}
\end{equation}
where ${\bm \xi}$ spans the parameter space.
The SNR and FIM for a detector network are $\rho^2=\sum_{\alpha=1}^{n}\rho^2_\alpha$ and $\Gamma_{ij} = \sum_{\alpha=1}^{n}\Gamma_{ij}^\alpha$.

For the inspiral signal, ${\bm \xi}=\{M_z$, $\eta$, $d_L$, $\theta_{s}$, $\varphi_s$, $\psi_s$, $\iota$, $t_c$, $\phi_0\}$,
where $M_z=M(1+z)$ is the redshifted total mass, $M=m_1+m_2$ is the total mass, $\eta=q/(1+q)^2$ is the symmetric mass ratio, $q=m_2/m_1\ge1$ is the mass ratio, $d_L$ is the luminosity distance.
For cosmological parameters, we use the Planck 2018 results: the Hubble constant $H_0=69.66$ km/s/Mpc, the matter energy density parameter $\Omega_m=0.3111$ and the cosmological constant $\Omega_\Lambda=0.6889$ \cite{Planck:2018vyg}.
It is hard to control the noise of space-based GW detectors below the frequency $\sim2\times10^{-5}$ Hz \cite{Baibhav:2020tma}, 
so we take $2\times10^{-5}$ Hz as the lower cutoff frequency.
We set $f_{\rm in}=\max\left(f_0, 2\times10^{-5}\ {\rm Hz}\right)$,
where $f_0$ is the frequency of $h_{2,-1}$ at one year before the coalescence given by Eq. \eqref{EqTime},
and set $f_{\rm out}$ to be the final frequency of $h_{4,-4}$ in the inspiral stage given by \textsc{imrphenomxhm}.

For mergers of nonspinning binary BHs, the spin of the remnant BH is only determined by the mass ratio $q$.
Thus, for the ringdown signal, ${\bm \xi}=\{q$, $M_z$, $d_L$, $\theta_s$, $\varphi_s$, $\psi_s$, $\iota$, $\phi_{220}$, $\phi_{330}$, $\phi_{210}$, $\phi_{440}\}$,
where $\phi_{\ell mn}$ is the phase of the corresponding quasinormal mode $(\ell,m,n)$ and $n$ is the overtone index.
We set $f_{\rm in}=\max\left(0.5f_{210}, 2\times10^{-5}\ {\rm Hz}\right)$ and $f_{\rm out}=2f_{440}$.

The covariance matrix of these parameters is
\begin{equation}
	\sigma_{ij}=\left\langle\Delta\xi^i\Delta\xi^j\right\rangle\approx (\Gamma^{-1})_{ij}.
\end{equation}
The angular uncertainty of the sky localization is evaluated as
\begin{equation}
	\Delta \Omega_s\equiv2\pi\sin\theta_s
	\sqrt{\sigma_{\theta_s\theta_s}\sigma_{\varphi_s\varphi_s}-\sigma^2_{\theta_s\varphi_s}},
\end{equation}
so the probability that the source lies outside an error ellipse enclosing the solid angle $\Delta\Omega$ is simply $e^{-\Delta\Omega/\Delta\Omega_s}$.

In this paper, we take $q=2$ and $t_c=0$.
For each binary with the specific total mass and redshift, we use Monte Carlo simulation to generate $10^3$ sources with $\{\cos\theta_s,\cos\iota\}$ uniformly distributed in $[-1, 1]$ and $\{\varphi_s,\psi_s,\phi_0,\phi_{\ell m n}\}$ uniformly distributed in $[0, 2\pi]$,
and obtain the median error of the parameter estimation and source localization.

\subsection{Characteristic strains}

\begin{figure}
	\includegraphics[width=0.95\columnwidth]{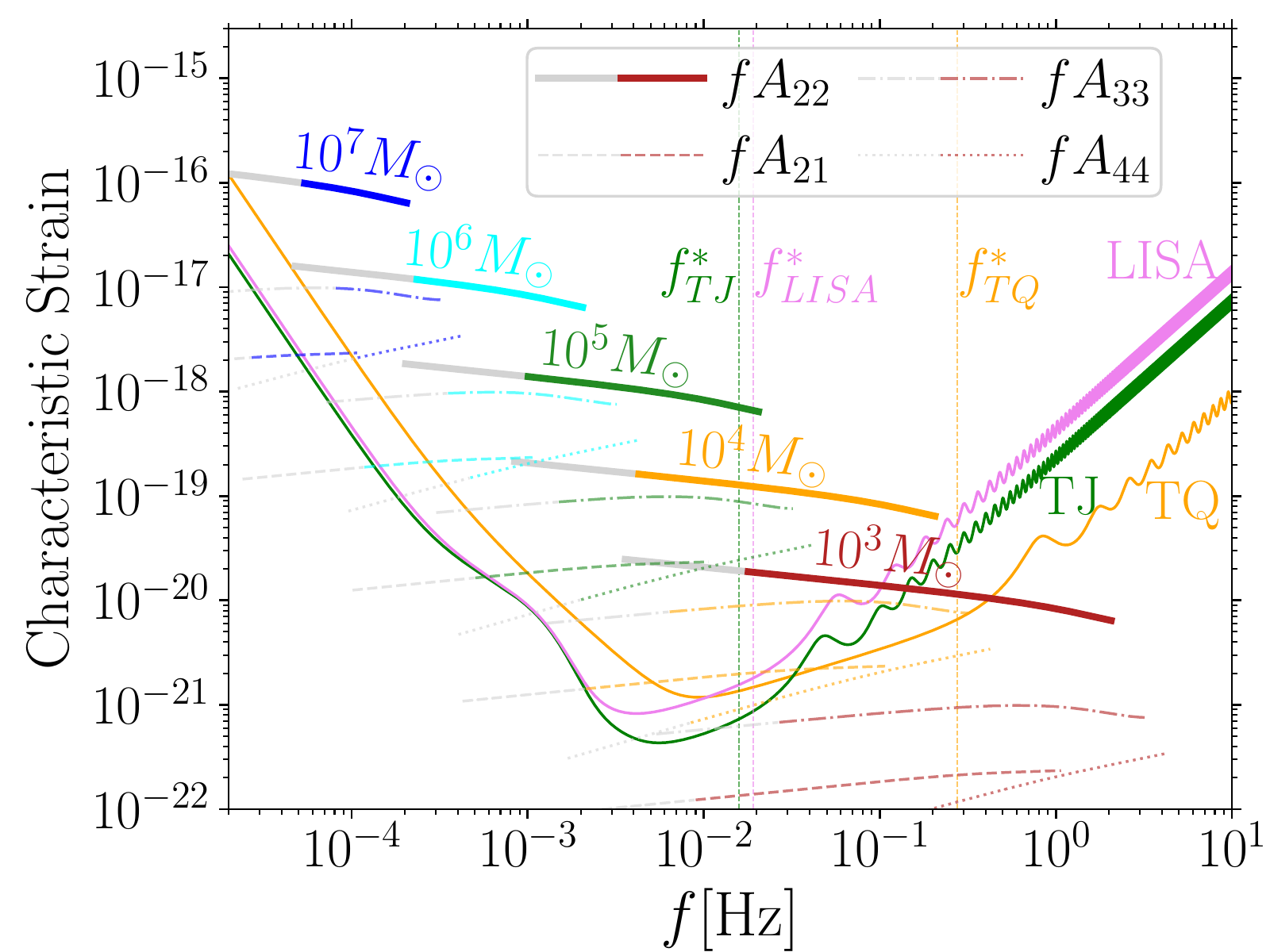}
	\caption{Characteristic strains for sensitivity curves of different detectors and various binaries.
    The gray-colored lines denote the inspiral signals from binaries with different total masses at $z = 1$,
	the line patterns (solid, dashed, dash-dotted, dotted) represent different harmonic modes,
	the first part (gray part) represents the inspiral signal from one year to 5 days before the coalescence, and the second part (colored part) represents the inspiral signal in the last 5 days.
	The vertical dotted lines are the transfer frequencies of LISA, TianQin, and Taiji.
	}
	\label{FigSn}	
\end{figure}

For inspiral signals, we consider the observation time of one year 
with the signal starting one year before the coalescence.
Figure \ref{FigSn} shows the sensitivity curves and the inspiral signals from various sources,
the gray-colored lines denote the inspiral signals from binaries with different total masses at $z = 1$,
the line patterns (solid, dashed, dash-dotted, dotted) represent different harmonic modes,
the first part (gray part) represents the inspiral signal from one year to 5 days before the coalescence, and the second part (colored part) represents the inspiral signal in the last 5 days.
From Fig. \ref{FigSn}, we see that the GW signal of the first part varies slowly with time,
while the GW signal of the second part varies rapidly.
Moreover, for the binary with $M=10^3 M_\odot$, 
the most contribution to SNR comes from the inspiral signal of the first part,
while for the binary with $M>10^4 M_\odot$,
the most contribution to SNR comes from the inspiral signal of the second part.

\section{Source localization}
\label{SecLocalization}

To investigate the effects of various factors on source localization, we construct two TianQin-like detectors, TQ1 and TQ2, with the same arm length as TianQin but pointing to $(\theta,\varphi)=(0^\circ, 0^\circ)$ and $(\theta,\varphi)=(\gamma_n, 0^\circ)$ respectively.
We also construct another three fiducial detectors, TianQin10L, TQ10L1, and TQ10L2, similar as TianQin, TQ1, and TQ2, respectively,
but with longer arm length $L=\sqrt{3}\times10^9$ m and longer rotation period.
All these constructed geocentric detectors orbit the Earth and further rotate around the Sun together with the Earth, without experiencing the Earth's own rotation.
Note we set the noise curves of these constructed geocentric detectors as the same as that of TianQin, which means TianQin10L, TQ10L1, and TQ10L2 have the arm length $L=\sqrt{3}\times10^9$ m only in the transfer function but not in the noise curve $P_n (f)$.

For the heliocentric detector, the normal vector of the detector plane spans a circular cone in one year with $60^\circ$ between the cone surface and the $z$ axis of the ecliptic plane.
LISA and Taiji point to the same direction when they arrive at the same location, which is determined by their orbital design.
For example, if LISA or Taiji is at $(\theta,\varphi)=(\pi/2,\varphi_0)$,
then the normal vector of its detector plane will point to $(\theta,\varphi)=(\pi/3,\varphi_0+\pi)$.
Since the separation angle $\varphi_{\rm sep}$ between LISA and Taiji is $40^\circ$, the angle between the normal vectors of their detector planes is $\gamma_n=\arccos[(1+3\cos\varphi_{\rm sep})/4]=34.5^\circ$.

\subsection{The angle between normal vectors}

\begin{figure}
	\includegraphics[width=\columnwidth]{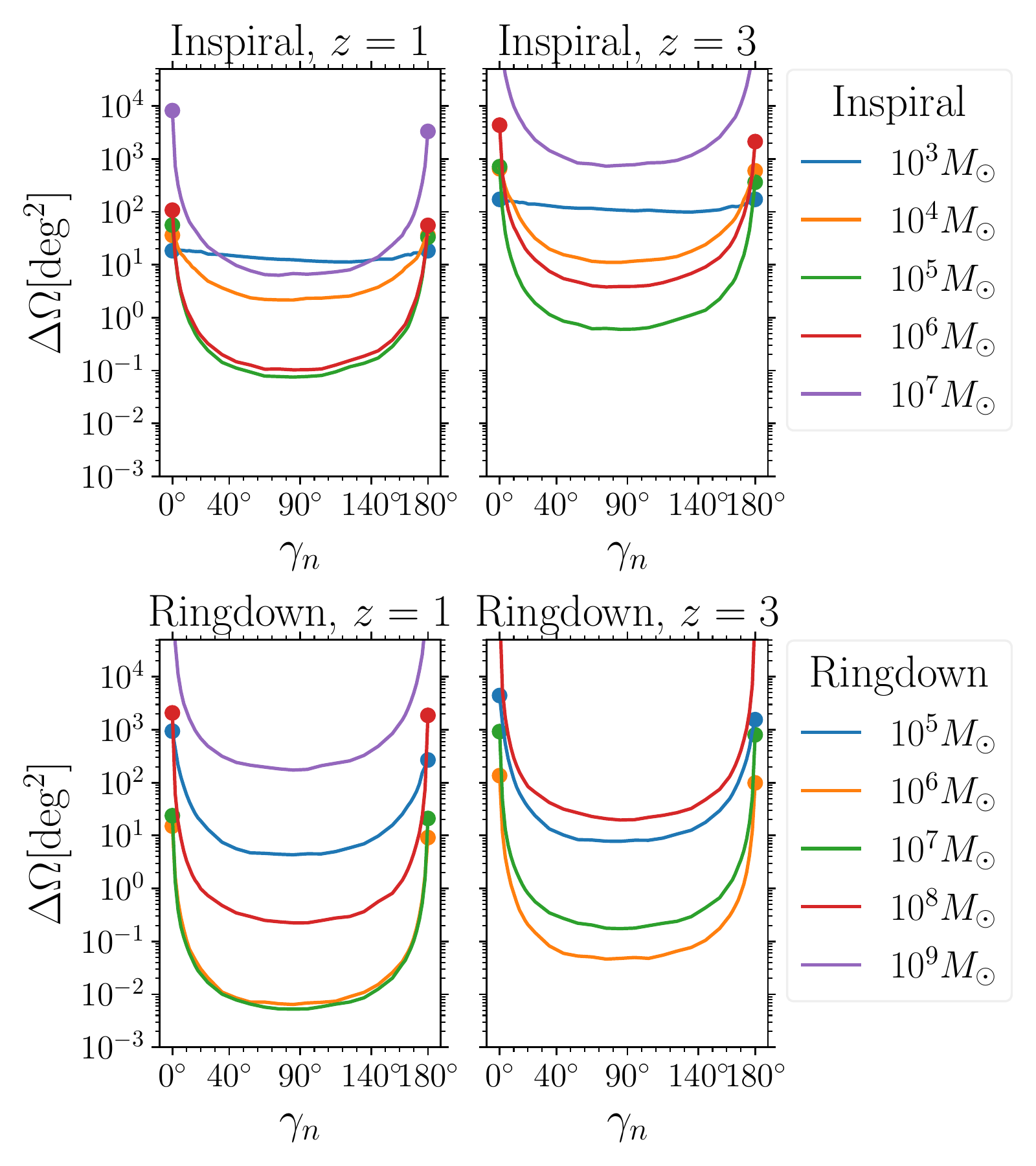}
	\caption{The median localization errors with the TQ1-TQ2 network for different angles between the normal vectors of the detector planes. The inspiral and ringdown signals are from binaries with different total masses at different redshifts.}
	\label{FigAngle}	
\end{figure}

\begin{figure*}
	\includegraphics[width=0.75\textwidth]{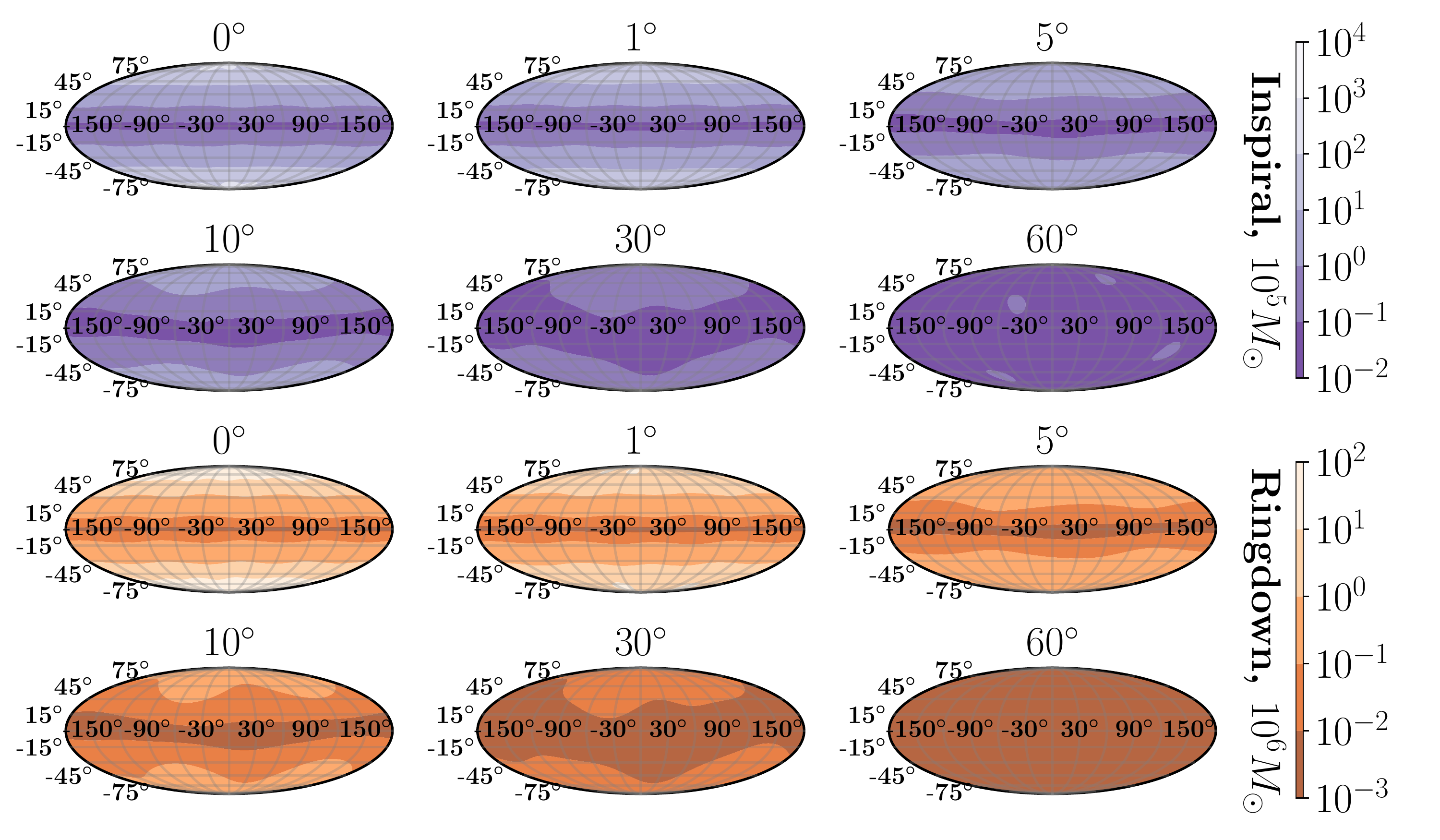}
	\caption{The sky dependence of the localization error for the TQ1-TQ2 network with different $\gamma_n$.
	The top two rows is for the inspiral signals from binaries with $M=10^5\ M_\odot$.
	The bottom two rows is for the ringdown signals from binaries with $M=10^6\ M_\odot$.}
	\label{FigSkyGamma}
\end{figure*}

Figure \ref{FigAngle} shows the median localization errors with the TQ1-TQ2 network for different angles between the normal vectors of the detector planes with inspiral signals and ringdown signals from binaries with different total masses at different redshifts.
From Fig. \ref{FigAngle},
we see that as the angle between the normal vectors increases,
the median error of the source localization decreases rapidly.
The network of TianQin-like detectors has the best angular resolution when the angle between the normal vectors of their detector planes is in the range $40^\circ-140^\circ$.
The network with $\gamma_n=180^\circ$ has a little better angular resolution than the network with $\gamma_n=0^\circ$,
because when $\gamma_n=0^\circ$ we set the three spacecrafts of TQ1 overlapping with those of TQ2 all the time.
The improvement to source localization by the network with $40^\circ\le\gamma_n\le140^\circ$ increases as the total mass of the binary increases,
and the improvement is negligible for binaries with $M=10^3\ M_\odot$.

Figure \ref{FigSkyGamma} shows the sky dependence of the localization error for the TQ1-TQ2 network with different angles between the normal vectors of the detector planes.
With the inspiral signal from the binary with $M=10^5\ M_\odot$ or the ringdown signal from the binary with $M=10^6\ M_\odot$,
as $\gamma_n$ increases, the localization errors with the TQ1-TQ2 network for signals from sources along the directions between the two detector planes decrease rapidly, and finally the skymap becomes uniform when $\gamma_n$ reaches $60^\circ$.

Since ringdown signals are transient (normally within one day),
the effect of the rotation of the detector is negligible in most cases.
For the ringdown signal, it can also use the response difference between different QNMs to localize the source.
If we combine two space-based detectors with $40^\circ\le \gamma_n\le140^\circ$,
the response difference between the two detectors is large enough in most cases,
which improves the source localization dramatically.
The angle between the normal vectors of the LISA-Taiji network is $34.5^\circ$,
and the angle between the normal vectors of the LISA-TianQin network or the Taiji-TianQin network varies from $34.7^\circ$ to $154.7^\circ$ periodically.
Thus, for ringdown signals, the LISA-Taiji network, the LISA-TianQin network, and the Taiji-TianQin network all improve the source localization dramatically compared with an individual detector, as shown in Ref. \cite{Zhang:2021kkh}.

\begin{figure*}
	\includegraphics[width=\textwidth]{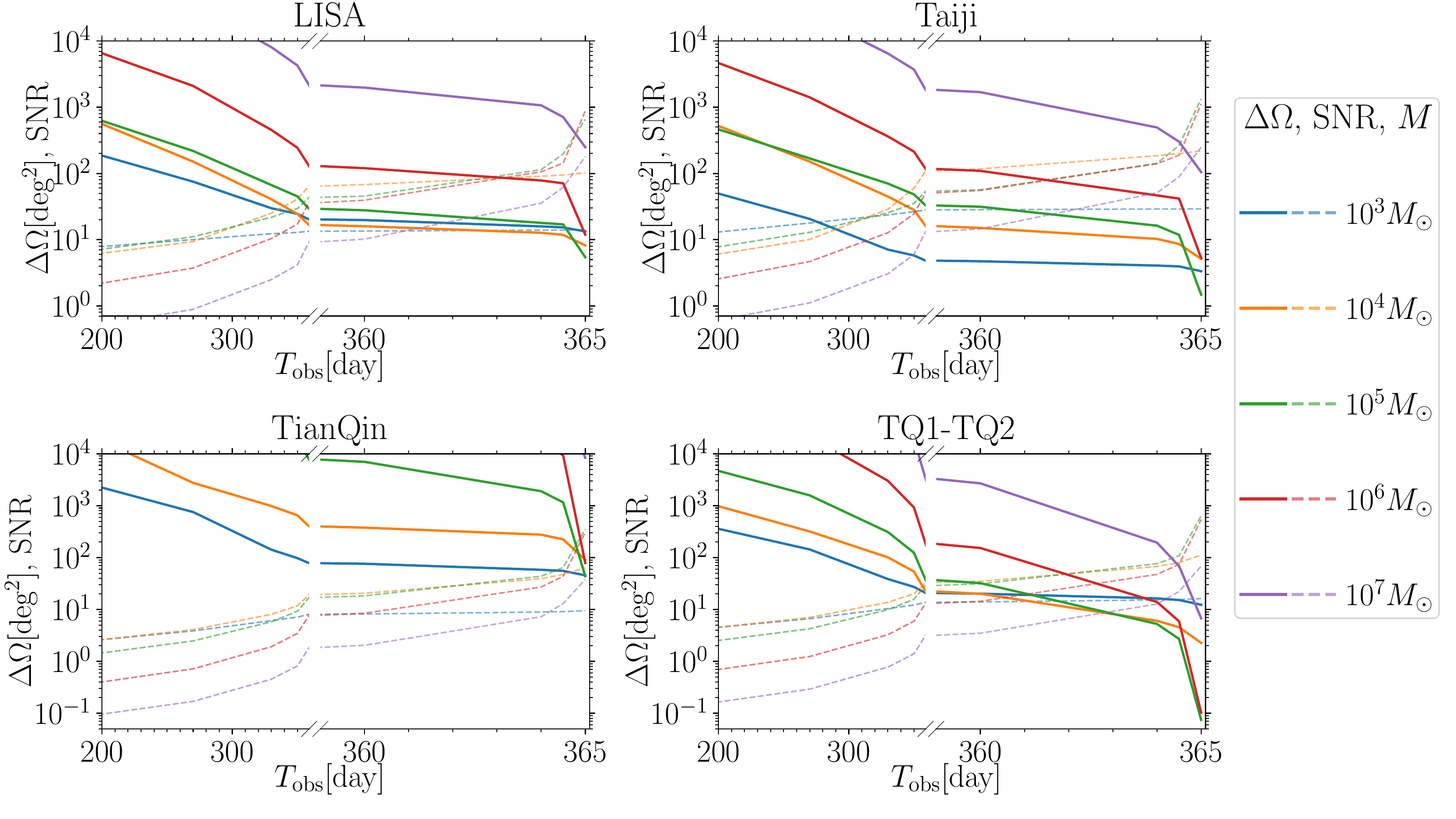}
	\caption{The median SNRs (dashed lines) and median localization errors (solid lines) with different detectors for different observation time starting from one year before the coalescence.
		For the TQ1-TQ2 network, $\gamma_n=90^\circ$.}
	\label{FigTime}
\end{figure*}

For inspiral signals, we consider the observation time of one year with the signal starting at one year before the coalescence.
Since the detector plane of geocentric detectors is fixed all the time,
it is easy to understand the dramatic improvement to the source localization when the detectors are combined.
Thus our discussion focuses on heliocentric detectors.

Figure \ref{FigTime} shows the median SNRs and median localization errors with different detectors for different observation times starting from one year before the coalescence.
From Fig. \ref{FigTime}, in the final few days, the median SNRs increase by a few times for binaries with $M=10^4\ M_\odot$, and by more than 1 or even 1.5 orders of magnitude for binaries with $M\ge10^5\ M_\odot$, but little for binaries with $M=10^3\ M_\odot$.

The inspiral signal of the first part varies slowly and lasts 360 days, which is the same as the continuous signal.
The heliocentric detector can localize continuous sources by using the amplitude modulation due to the varying orientation of the detector plane and the phase modulation due to the orbital motion of the detector center around the Sun,
and the geocentric detector with the fixed detector plane can localize continuous sources only by using the phase modulation \cite{Zhang:2020hyx}.
The significant difference between the source localization of the heliocentric detector and the geocentric detector is mainly caused by the amplitude modulation.
The analysis of source localization for monochromatic waves shows that the heliocentric detector has much better localization accuracy than the geocentric detector due to the amplitude modulation in the case of $f<10^{-3}$ Hz, and has similar localization accuracy as the geocentric detector in the case of $f>10^{-3}$ Hz because the phase modulation becomes dominant \cite{Cutler:1997ta,Blaut:2011zz,Zhang:2020drf}.
Thus, for the observation of the first part, the heliocentric detector can be regarded as a network of detectors with different orientations (due to amplitude modulation) and it has much better localization accuracy than the geocentric detector in the case of $M\ge10^4\ M_\odot$ (amplitude modulation dominates) and has similar localization accuracy as the geocentric detector in the case of $M=10^3\ M_\odot$ (phase modulation dominates).
As shown in Fig. \ref{FigTime}, at $T_{\text{obs}}=360$ days,
LISA has much better source localizations than TianQin in the case of $M\ge10^4\ M_\odot$,
and similar source localizations as TianQin in the case of $M=10^3\ M_\odot$,
similar source localizations as the TQ1-TQ2 network for binaries with $10^3\ M_\odot\le M\le 10^7\ M_\odot$.
Taiji has better localization accuracy than LISA due to its lower noise curve.

As seen from Fig. \ref{FigSn}, for the first part of the inspiral signal,
the GW frequencies for $10^3\ M_\odot$ and $10^4\ M_\odot$ are around $10^{-2}$ Hz and $2\times10^{-3}$ Hz respectively.
From Fig. \ref{FigTime}, we see that at $T_{\text{obs}}=360$ days,
for binaris with $M=10^3\ M_\odot$ and $M=10^4\ M_\odot$ the median SNRs and median localization errors with LISA are about $\{10,10\ \text{deg}^2\}$ and $\{70,10\ \text{deg}^2\}$, respectively;
the results with TianQin are $\{10,50\ \text{deg}^2\}$ and $\{20,500\ \text{deg}^2\}$, respectively.
Reference \cite{Zhang:2020hyx} employed the monochromatic wave model,
and find that in the case of $\rho=10$ the median localization errors with LISA for $10^{-2}$ Hz and $10^{-3}$ Hz are about $2.5\ \text{deg}^2$ and $150\ \text{deg}^2$, respectively;
the results with TianQin are about $3.5\ \text{deg}^2$ and $5000\ \text{deg}^2$, respectively.
From the relationship $\Delta\Omega\propto 1/\rho^2$ for the monochromatic wave model, the localization accuracy of the detector with the first part of the inspiral signal is a few times worse than that given by the monochromatic wave model.

The inspiral signal of the second part varies fast and lasts 5 days, which is the same as the transient signal.
The second part is the key to the improvement to the source localization by the network.
For binaries with $M>10^4 M_\odot$,
the most contribution to SNR comes from the inspiral signal in the final 5 days,
when the motion of the heliocentric detector is ignorable.
Thus, in this case, a single heliocentric detector can only be regarded as a network of detectors with very small $\gamma_n$, which has little effect in improving source localizations.
Although in the case of $M>10^4\ M_\odot$ the SNR of the second part is larger than the SNR of the first part by 1 or even 1.5 orders of magnitude,
it only decreases the localization errors with a single heliocentric detector by one order of magnitude at most.
As seen from Fig. \ref{FigTime}, with the second part of the signal, the TQ1-TQ2 network can decrease the median localization errors by one order of magnitude for $M=10^4\ M_\odot$ and by even three orders of magnitude for $M=10^6\ M_\odot$ compared with the first part of the signal.
For $M=\{10^4, 10^5, 10^6, 10^7\}\ M_\odot$,
the localization errors with the TQ1-TQ2 network are smaller than those with LISA or Taiji by $\{2, 20, 50, 20\}$ times.
Note that these detectors have the same order of magnitude for SNRs,
so the dramatic improvement on sky localization by the TQ1-TQ2 network is due to the large $\gamma_n$ in the network.
For $M<10^4\ M_\odot$, the SNR of the second part is equal to or smaller than that of the first part,
thus in this case the improvement by the network is negligible.
The more SNR contributed by smaller parts in a shorter time,
the better improvement by the network.
Thus, the detector network can dramatically improve source localizations for short GW sources and long sources with most contributions to SNR coming from a small part of the signal in a short time.
A single space-based detector can localize sources even with transient ringdown signals due to the large SNRs and a pair of interferometers in it, as shown in Figs. \ref{FigAngle} and \ref{FigSkyDet} and Refs. \cite{Baibhav:2020tma,Zhang:2021kkh},
thus the dramatic improvement by the network is not because a single space-based detector cannot localize sources with inspiral signals of the second part.

Ground-based detectors have observed numerous compact binary coalescences, but have not observed continuous sources yet.
These transient signals are the same as the ringdown signals and the inspiral signals from binaries with $M\ge10^5\ M_\odot$.
For the Hanford-Livingston network, the angle between the normal vectors of LIGO Hanford and LIGO Livingston is $27^\circ$, and the typical SNR and localization error are about 10 and 1500 ${\rm deg}^2$ \cite{LIGOScientific:2018mvr,LIGOScientific:2020ibl}.
TianQin has two interferometers, which is also a network with $\gamma_n=0^\circ$.
For TianQin, the median SNRs and median localization errors of inspiral signals from binaries with $M=\{10^5,10^6,10^7\}\ M_\odot$ at $z=1$ are about \{350, 300, 36\} and \{40, 70, 8000\} ${\rm deg}^2$;
the median values of ringdown signals from binaries with $M=\{10^5,10^6,10^7,10^8\}\ M_\odot$ at $z=1$ are about \{70, 2000, 2000, 300\} and \{900, 15, 25, 2000\} ${\rm deg}^2$.
With large SNRs, TianQin has a smaller typical localization error than the Hanford-Livingston network.

For the LIGO-Virgo network, the angle between the normal vectors of LIGO and Virgo is $79^\circ$, and the typical SNR and localization error are about 12 and 400 ${\rm deg}^2$ \cite{LIGOScientific:2018mvr,LIGOScientific:2020ibl}.
For the TQ1-TQ2 network with $\gamma=90^\circ$, the median SNRs and median localization errors of inspiral signals from binaries with $M=\{10^5,10^6,10^7\}\ M_\odot$ at $z=1$ are about \{600, 500, 65\} and \{0.08, 0.1, 7\} ${\rm deg}^2$;
the median values of ringdown signals from binaries with $M=\{10^5,10^6,10^7,10^8\}\ M_\odot$ at $z=1$ are about \{90, 2300, 2500, 400\} and \{4, 0.006, 0.005, 0.2\} ${\rm deg}^2$.
The TQ1-TQ2 network with four interferometers and without time delay has a much smaller typical localization error than the LIGO-Virgo network with three interferometers and time delay due to the large SNRs.
If we only use two interferometers of the TQ1-TQ2 network, the localization accuracy would become about 6 times worse for inspiral signals from binaries with $M\ge10^5\ M_\odot$ and about 60 times worse for ringdown signals.

\subsection{The time delay}

\begin{figure}
	\includegraphics[width=0.95\columnwidth]{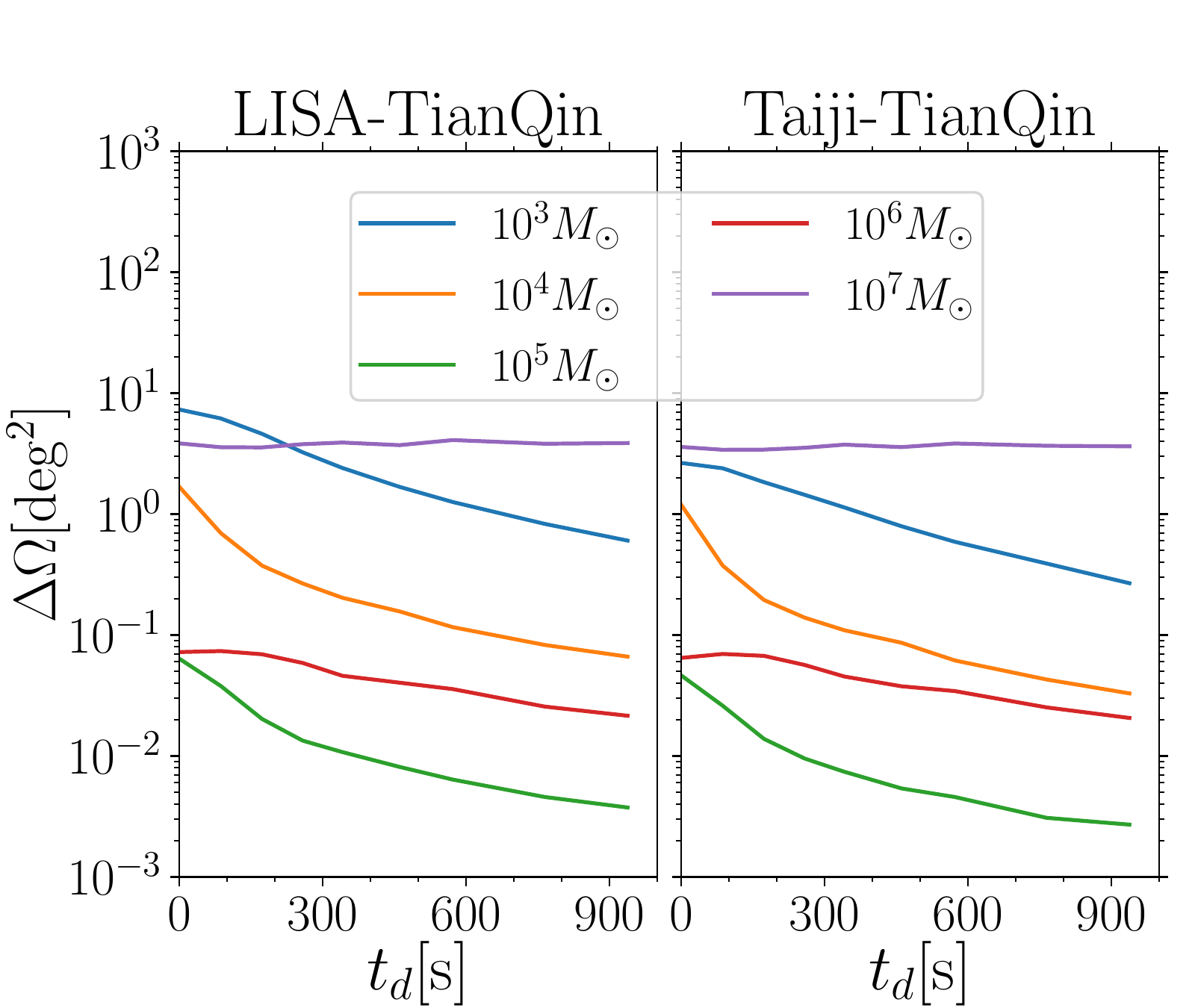}
	\caption{The median localization errors with the LISA-TianQin network (left) and the Taiji-TianQin network (right) for different time delay with the inspiral signals from binaries with different total masses at $z=1$.
	}
	\label{FigDelay}	
\end{figure}

The time delay between detectors in the network contains information about the source position, which can help to localize the source.
We use $t_d=L_d/c$ to represent the time delay, where $L_d$ is the distance between two detectors.

Figure \ref{FigDelay} shows the median localization errors with the LISA-TianQin network and the Taiji-TianQin network for different time delays with inspiral signals from binaries with different total masses at $z=1$.
When the observation starts, we set LISA pointing to $(\theta,\varphi)=(\pi/3,\pi)$ for the LISA-TianQin network, and set Taiji pointing to $(\theta,\varphi)=(\pi/3,\pi)$ for the Taiji-TianQin network, so that $\gamma_n=66.6^\circ$ for the two networks at the coalescence time.
Then, we change the location of TianQin to make different time delays.

\begin{table}
	\centering	
	\resizebox{0.9\columnwidth}{!}{
		\begin{tabular}{|c|c|c|c|c|c|}
			\hline
			$M(M_\odot)$& LISA & Taiji & TianQin & TQ1-TQ2 & LISA-TianQin\\
			\hline
			$10^3$& 13.2 & 3.34 & 45.5 & 12.3 & 2.40  \\
			\hline
			$10^4$& 8.19 & 5.13 & 90.6 & 2.27 &  0.203 \\
			\hline
			$10^5$& 5.42 & 1.48 & 43.7 & 0.075 &  0.011 \\
			\hline
			$10^6$& 11.9 & 5.30 & 78.7 & 0.103 & 0.046 \\
			\hline
			$10^7$& 248.5 & 104.9 & 8367 & 6.77 & 3.90 \\
			\hline
		\end{tabular}
	}
	\caption{Median localization errors with different detectors for inspiral signals from binaries with different total masses at $z=1$.
	For the TQ1-TQ2 network, the angle between normal vectors is $90^\circ$.
	For the LISA-TianQin network, the time delay is 340 s, and the angle between the normal vectors is $66.6^\circ$ at the coalescence time.}
	\label{TabInsp}
\end{table}

Table \ref{TabInsp} shows the median localization errors with different detectors for inspiral signals from binaries with different total masses at $z=1$,
As seen from Fig. \ref{FigDelay} and Table \ref{TabInsp},
in the case of $t_d=0$ s,
the results with the LISA-TianQin network and the Taiji-TianQin network are similar to those with the TQ1-TQ2 network;
in the case of $t_d=340$ s,
the results with the LISA-TianQin network and the Taiji-TianQin network are better than those with the TQ1-TQ2 network by a few times for $M=10^6\ M_\odot$ and by one order of magnitude for $10^3\ M_\odot\le M \le 10^5\ M_\odot$.
Moreover, for $M\ge10^7\ M_\odot$,
the contribution of the time delay to the source localization is negligible,
because the GW frequency is lower than $1/(2\text{AU})=10^{-3}$ Hz, where $2\text{AU}$ is the diameter of the heliocentric orbit.

For the equal-mass binaries with $M=10^5\ M_\odot$ at $z=1$, the localization accuracy of the LISA-Taiji network is improved by two orders of magnitude as the separation angle between LISA and Taiji varies from $3^\circ$ to $40^\circ$,
and by only 5 times as separation angle varies from $40^\circ$ to $180^\circ$ \cite{Ruan:2020smc}.
The dramatic improvement as $\varphi_{\rm sep}$ varies from $3^\circ$ to $40^\circ$ is mainly because $\gamma_n$ varies from a few degrees to $34.5^\circ$,
and partly because $t_d$ varies from 26 s to 340 s.
The less improvement as $\varphi_{\rm sep}$ varies from $40^\circ$ to $180^\circ$ is only caused by the time delay as shown in Fig. \ref{FigDelay}, and there is almost no improvement as $\gamma_n$ varies from $34.5^\circ$ to $120^\circ$.
For ground-based detectors, since $t_d=0.01$ s for the Hanford-Livingston network and $t_d=0.027$ s for the LIGO-Virgo network,
the ground-based detector network can improve the source localization for sources with $f\ge37$ Hz.
With the timing triangle method, Ref. \cite{Fairhurst:2009tc} shows that the localization accuracy of the LIGO-Virgo network is about 3 times better than that for the Hanford-Livingston network,
which is consistent with the difference between $t_d=300$ s and $t_d=100$ s in Fig. \ref{FigDelay}.

\subsection{The higher harmonics}

Table \ref{TabInspH22} shows median localization errors with different detectors for only the $(2, 2)$ mode of inspiral signals from binaries with different total masses at $z=1$.
From Tables \ref{TabInsp} and \ref{TabInspH22}, for LISA and Taiji, the improvement by higher harmonics is negligible in the case of $M\le 10^4\ M_\odot$,
is about 35\% in the case of $M=10^5\ M_\odot$,
and is about 7 times in the case of $10^6\ M_\odot\le M \le 10^7\ M_\odot$;
for the detector network, the improvement by higher harmonics is negligible in the case of $M\le 10^6\ M_\odot$, and is only 60\% in the case of $M=10^7\ M_\odot$;
for TianQin, the improvement by higher harmonics is one or two orders of magnitudes in the case of $M\ge 10^5\ M_\odot$.
In the ringdown stage, a single space-based detector cannot localize sources without higher harmonics, but the localization accuracy of the TQ1-TQ2 network varies little without higher harmonics.
Thus, higher harmonics are unimportant for source localizations with the detector network all the time,
are necessary for source localizations with a single space-based detector in the ringdown stage,
are important for source localizations with heliocentric detectors in the case of $M\ge 10^6\ M_\odot$ in the inspiral stage,
and are very important for source localizations of geocentric detectors in the case of $M\ge 10^5\ M_\odot$ in the inspiral stage.

\begin{table}
	\centering	
	\resizebox{0.9\columnwidth}{!}{
		\begin{tabular}{|c|c|c|c|c|c|}
			\hline
			$M(M_\odot)$& LISA & Taiji & TianQin & TQ1-TQ2 & LISA-TianQin\\
			\hline
			$10^3$& 14.2 & 3.82 & 65.4 & 15.3 & 2.50  \\
			\hline
			$10^4$& 8.38 & 5.26 & 171.3 & 2.28 &  0.205 \\
			\hline
			$10^5$& 7.20 & 2.17 & 2691 & 0.077 &  0.011 \\
			\hline
			$10^6$& 61.0 & 39.4 & $>10^5$ & 0.116 & 0.058 \\
			\hline
			$10^7$& 1193 & 730 & $>10^5$ & 9.22 & 6.04 \\
			\hline
		\end{tabular}
	}
	\caption{Median localization errors with different detectors for only the $(2, 2)$ mode of inspiral signals from binaries with different total masses at $z=1$.
	For the TQ1-TQ2 network, the angle between normal vectors is $90^\circ$.
	For the LISA-TianQin network, the time delay is 340 s, and the angle between the normal vectors is $66.6^\circ$ at the coalescence time.}
	\label{TabInspH22}
\end{table}

\begin{figure*}
	\includegraphics[width=\textwidth]{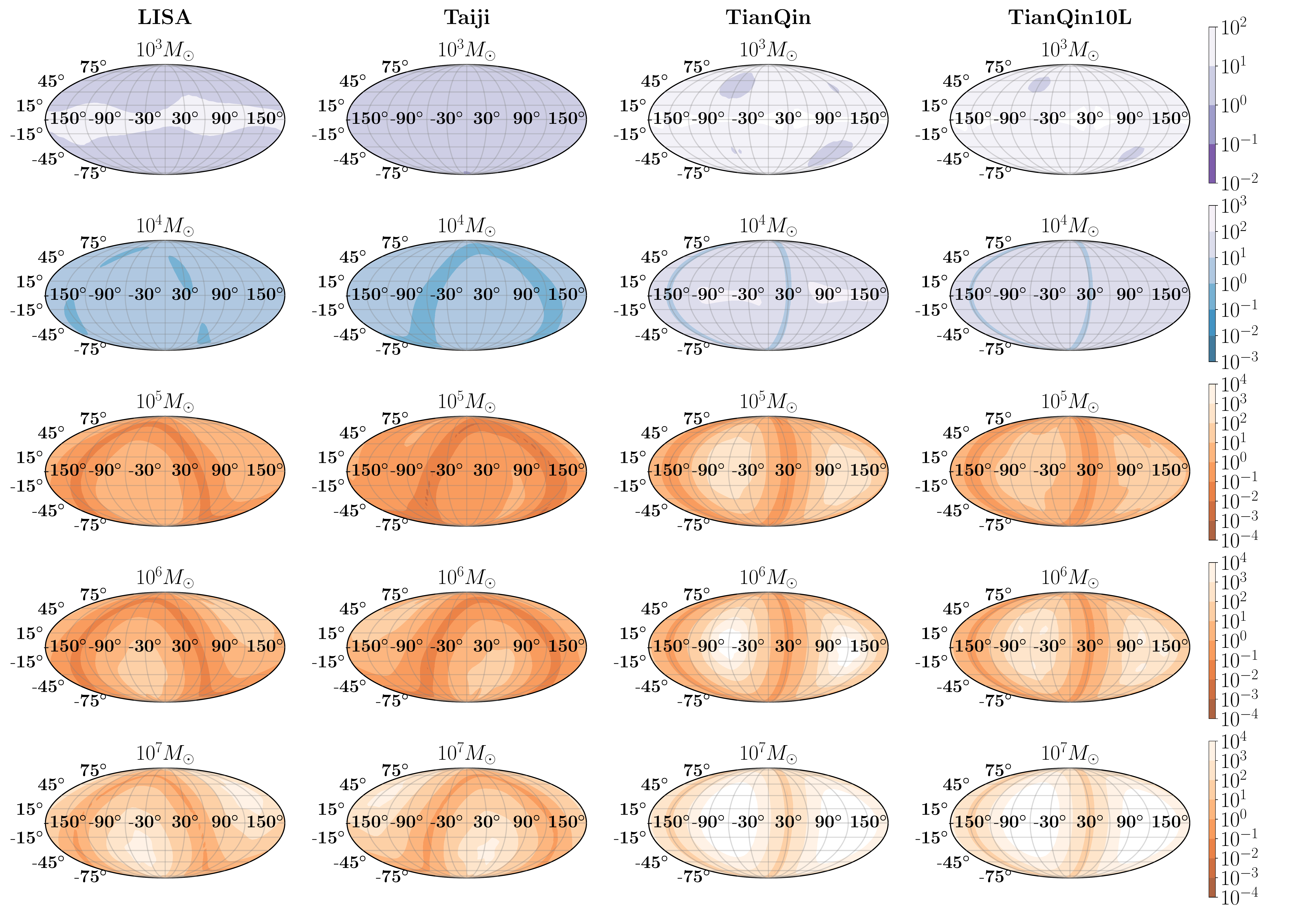}
	\caption{The sky dependence of localization errors for LISA, TianQin, TianQin10L, and Taiji with inspiral signals from binaries with different total masses at $z=1$.
	When the observation starts or the coalescence happens, we set LISA pointing to $(\theta,\varphi)=(\pi/3,8\pi/9)$, and Taiji pointing to $(\theta,\varphi)=(\pi/3,10\pi/9)$.
	TianQin and TianQin10L point to $(\theta,\varphi)= (94.7^\circ, 120.5^\circ)$ all the time.}
	\label{FigSkyDet}	
\end{figure*}

Figure \ref{FigSkyDet} shows the sky dependence of the localization errors for LISA, Taiji, TianQin, and TianQin10L with inspiral signals from binaries with different total masses at $z=1$.
When the observation starts or the coalescence happens, we set LISA pointing to $(\theta,\varphi)=(\pi/3,8\pi/9)$, and Taiji pointing to $(\theta,\varphi)=(\pi/3,10\pi/9)$.
From Fig. \ref{FigSkyDet}, we see that all the detectors have the worst angular resolution for sources along the equator plane in the case of $M=10^3\ M_\odot$, and have the best angular resolution for sources along the detector plane in the case of $M\ge10^4\ M_\odot$.
In the case of $M=10^3\ M_\odot$,
the most contribution to SNR comes from the inspiral signal in the first 360 days.
For LISA and Taiji, the amplitude modulation makes the skymap more uniform and the phase modulation makes the localization accuracy worse for sources along the equator plane;
For TianQin, there is only the phase modulation which makes the localization accuracy worse for sources along the equator plane \cite{Zhang:2020hyx}.
In the case of $M\ge10^4\ M_\odot$,
the most contribution to SNR comes from the inspiral signal in the final 5 days,
when the heliocentric detector can be modeled as a network with small $\gamma_n$ with the best angular resolution for sources along the detector plane.
For all detectors, higher harmonics make the localization accuracy better for sources along the detector plane (similar to the results of ringdown signals in Refs. \cite{Zhang:2021kkh,Baibhav:2020tma}).

\begin{figure}
	\includegraphics[width=0.95\columnwidth]{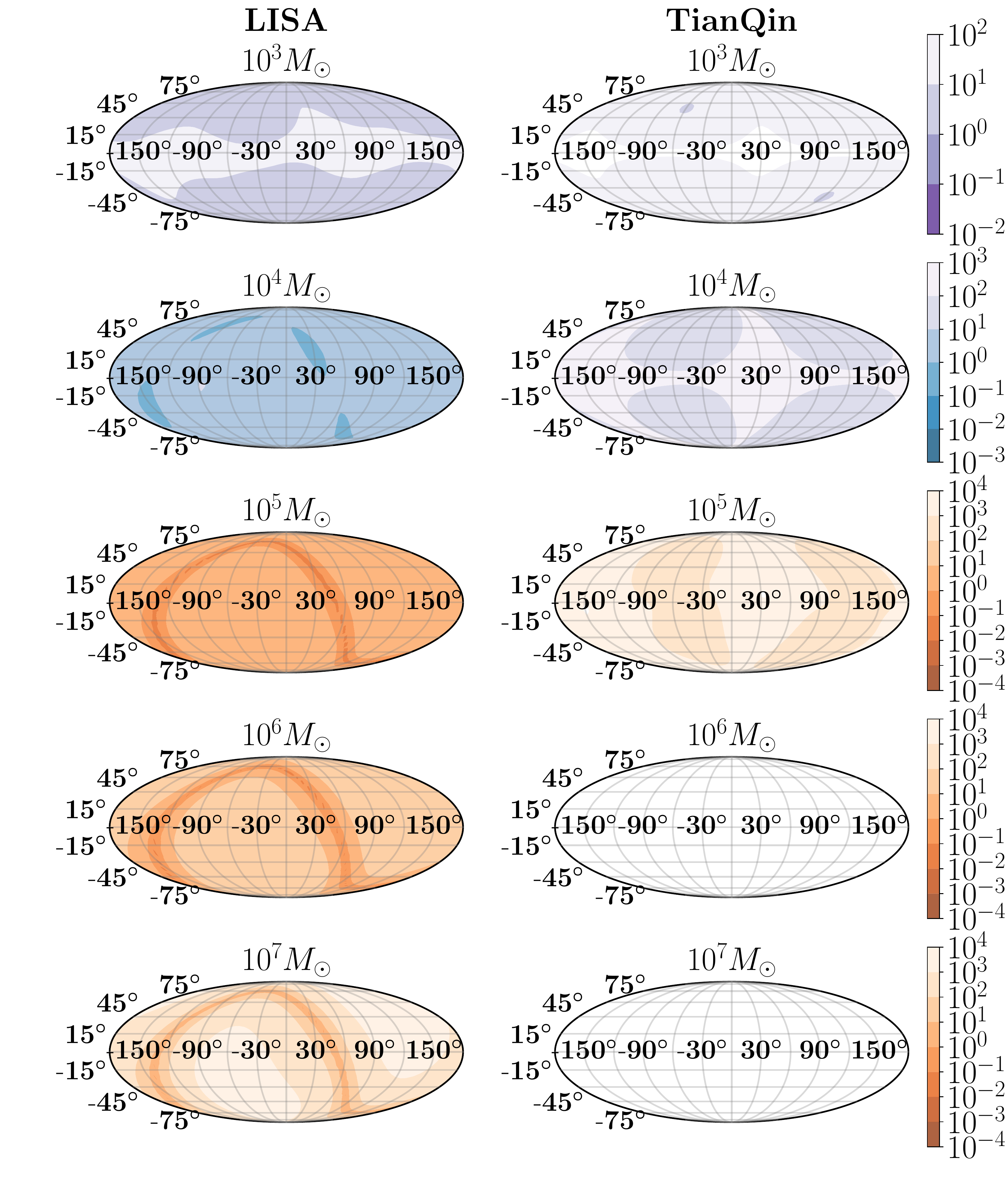}
	\caption{The sky dependence of localization errors with LISA and TianQin for only the $(2,2)$ mode of inspiral signals from binaries with different total masses at $z=1$.
	}
	\label{FigSkyH22}	
\end{figure}

Figure \ref{FigSkyH22} shows the sky dependence of the localization errors for LISA and TianQin with only the $(2,2)$ mode of inspiral signals from binaries with different total masses at $z=1$.
In the case of $M\ge10^4\ M_\odot$, although LISA can be modeled as a network with small $\gamma_n$, which makes the angular resolution better for sources along the detector plane,
the $\gamma_n$ is not large enough to maintain the localization accuracy as the detector networks.
Since the detector plane of the geocentric detector is fixed, the localization accuracy of TianQin becomes much worse if without higher harmonics.
With only the $(2, 2)$ mode, TianQin has the worst localization accuracy for sources along the detector plane, because the tensor response function reaches the minimum \cite{Liang:2019pry}.
If we rotate the detector plane of TianQin slowly, it will also get the best angular resolution for sources along the detector plane at the coalescence time.

Reference \cite{Baibhav:2020tma} shows that the higher harmonics with different dependence on inclination angle are necessary for source localizations with LISA.
The sky dependence of localization errors for space-based detectors with inspiral signals from binaries with $M\ge10^4\ M_\odot$ (second part dominates) is consistent with the sky dependence of localization errors for space-based detectors with ringdown signals as shown in Refs. \cite{Baibhav:2020tma,Zhang:2021kkh}.

\subsection{The transfer function}

Since the effect of the rotation period of spacecrafts is small \cite{Zhang:2020hyx},
the resulting difference between TianQin and TianQin10L can only be caused by the transfer function, which contains the information about the source position as shown in Eq. \eqref{EqTransfer}.
For GWs with $0.5f^*\le f \le 5f^*$,
$\cal{T}$ slightly weakens the response and dramatically improves the source localization.
From Fig. \ref{FigSn}, we see that the transfer functions of LISA, TianQin10L,
and Taiji can improve source localizations for inspiral signals from sources with $M \le 10^6\ M_\odot$ at $z=1$,
the transfer functions of TianQin can improve source localizations for inspiral signals from sources with $M \le 10^5\ M_\odot$ at $z=1$.
As seen from Fig. \ref{FigSkyDet}, 
TianQin10L has better localization accuracy than TianQin for $10^4\ M_\odot\le M \le 10^6\ M_\odot$, and similar sky dependence of localization error as TianQin.

Reference \cite{Seto:2002uj} shows that the transfer function of LISA can improve the localization accuracy by ten times for binaries with $M=10^5\ M_\odot$ and by a few times for binaries with $M=\{10^4,10^6\}\ M_\odot$, which is consistent with the range $0.5f^*\le f \le 5f^*$.

\begin{figure}
	\includegraphics[width=0.95\columnwidth]{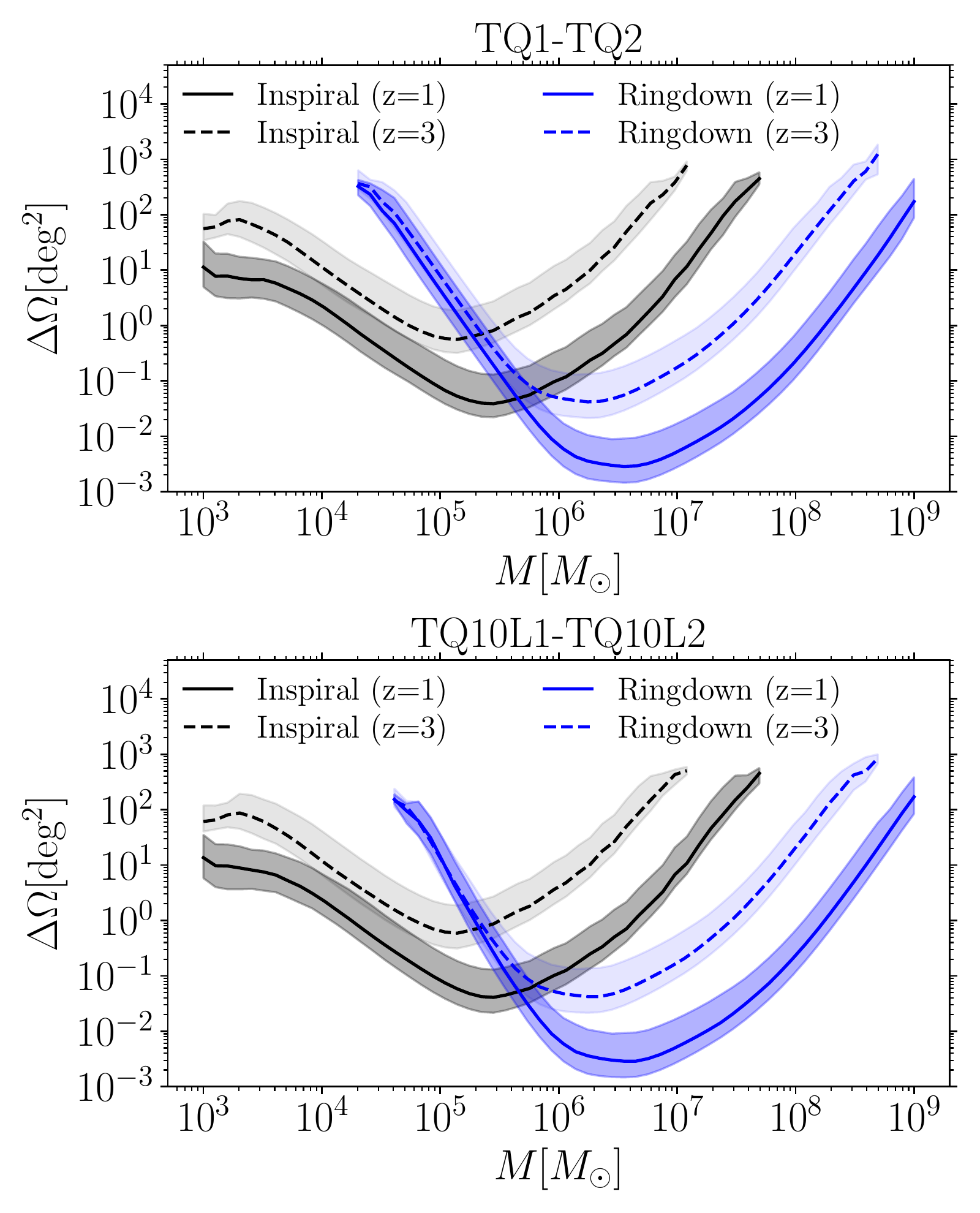}
	\caption{The median localization errors and $1\sigma$ localization errors with the TQ1-TQ2 network and TQ10L1-TQ10L2 network for inspiral signals and ringdown signals from binaries with different total masses at different redshifts.
	Here we take $\gamma_n=90^\circ$.}
	\label{FigMass}	
\end{figure}

Figure \ref{FigMass} shows the best localization accuracy for the TQ1-TQ2 network and TQ10L1-TQ10L2 network without considering the time delay.
From Fig. \ref{FigMass}, the results of the two networks are very similar.
For both networks, the source localization with ringdown signals is much better than that with inspiral signals when the redshifted total mass of the binary is larger than $1.5\times10^6\ M_\odot$.
Thus, the effect of the transfer function is negligible in the network with $40^\circ\le\gamma_n\le140^\circ$.

\section{Conclusion}
\label{SecConclusion}

The sky localization of GW sources is an important scientific objective for GW observations.
The network of space-based GW detectors dramatically improves the sky localization accuracy compared with an individual detector not only in the inspiral stage but also in the ringdown stage.
We find that the angle between the normal vectors of the detector planes dominates the improvement.
The detector network has the best angular resolution when the angle between the normal vectors is in the range $40^\circ-140^\circ$.
The angle between the normal vectors of the LISA-Taiji network is $34.5^\circ$,
and the angle between the normal vectors of the LISA-TianQin network or the Taiji-TianQin network varies from $34.7^\circ$ to $154.7^\circ$ periodically.
Thus, the LISA-Taiji network, the LISA-TianQin network, and the Taiji-TianQin network, all improve the source localization dramatically compared with an individual detector.
We also find that the detector network dramatically improves source localizations for short GW sources and long GW sources with most contributions to the SNR coming from a small part of the signal in a short time,
and the more SNR contributed by smaller parts, the better improvement by the network.
Furthermore, the improvement is more dramatic for heavier sources because a smaller part of the signal in a shorter time makes a major contribution to the SNR.
Note that a network of two detectors only has a little bigger SNR than a single detector because $\rho=\sqrt{\rho_1^2+\rho_2^2}\le\sqrt{2}\rho_2$ if $\rho_1\le\rho_2$,
so the dramatic improvement on sky localization by the network is not due to the increase of SNR but the large angle spanned between the normal vectors of the detector planes in the network.

The time delay can further decrease the localization error by a few times or even one order of magnitude,
but its effect is negligible for binaries with $M\ge10^7\ M_\odot$ in the inspiral stage.

Higher harmonics are unimportant for source localizations with the detector network all the time,
are necessary for source localizations with a single space-based detector in the ringdown stage,
are important for source localizations with heliocentric detectors in the case of $M\ge 10^6\ M_\odot$ in the inspiral stage,
and are very important for source localizations of geocentric detectors in the case of $M\ge 10^5\ M_\odot$ in the inspiral stage.

The transfer function also helps sky localization for GWs with $0.5f^*\le f \le 5f^*$,
thus the longer arm length helps to localize heavier sources.
However, the effect of the transfer function is negligible in the detector network with the angle between the normal vectors in the range $40^\circ-140^\circ$.

These results are helpful to improve the detector design and explore the scientific potential of space-based GW detectors.

\begin{acknowledgments}
	This research is supported in part by the National Key Research and Development Program of China under Grant No. 2020YFC2201504,
	the National Natural Science
	Foundation of China under Grant No. 11875136,
	and the Major Program of the National Natural Science Foundation of China under Grant No. 11690021.
\end{acknowledgments}


%

\end{document}